\documentclass[10pt,letterpaper]{article}
\usepackage{opex3,epstopdf,amsmath}

\begin{document}

\title{Rayleigh scattering boosted multi-GHz displacement sensitivity in whispering gallery opto-mechanical resonators}

\author{Siddharth Tallur and Sunil A. Bhave}

\address{OxideMEMS Lab, Cornell University,\\ Ithaca, NY 14853}

\email{sgt28@cornell.edu} 



\begin{abstract}
Finite photon lifetimes for light fields in an opto-mechanical cavity impose a bandwidth limit on displacement sensing at mechanical resonance frequencies beyond the loaded cavity photon decay rate. Opto-mechanical modulation efficiency can be enhanced via multi-GHz transduction techniques such as piezo-opto-mechanics at the cost of on-chip integration. In this paper, we present a novel high bandwidth displacement sense scheme employing Rayleigh scattering in photonic resonators. Using this technique in conjunction with on-chip electrostatic drive in silicon enables efficient modulation at frequencies up to 9.1GHz. Being independent of the drive mechanism, this scheme could readily be extended to piezo-opto-mechanical and all optical transduced systems.
\end{abstract}



\section{Introduction}

Cavity opto-mechanical systems have enabled a wide range of experiments pertaining to ultrahigh optical readout sensitivity, photon-phonon translation and mechanical signal amplification in the same device platform \cite{cho,detector,kippvahala,coolvibrations}. With a view towards realizing truly on-chip integration, several demonstrations of cavity opto-mechanical systems with on-chip electrodes for transduction of motion using electrostatic capactive actuation have been demonstrated \cite{ipj,painterelec,nitrideelec}. However, the advantage of employing action-at-a-distance electrostatic actuation comes with the drawback of the inherently low bandwidth of this transduction scheme. The transduction efficiency could be boosted via partial gap transduction using sub-100nm resonator-electrode gap spacings \cite{T13}, however this only enables frequency scaling up to a couple of GHz. Piezoelectric actuation on the other hand, has traditionally been used in micro-electro mechanical systems (MEMS) for beyond-GHz actuation \cite{otis,ruby,piazza}. Recently this scheme was adopted to opto-mechanical systems, and piezo-opto-mechanical systems operating at mechanical resonance frequencies in 3-4GHz range were demonstrated \cite{hongtang,cleland}. However, an integrated piezoelectric resonator necessitates metal electrodes atop the resonator for actuating motion \cite{piazza}. This would lead to huge optical loss in the metal and hence implementation of piezo-opto-mechanical systems till date has relied on employing electrodes located far away from the resonator ($\approx$50$\mu$m) so that the optical performance is not compromised \cite{hongtang,cleland}. 

It is thereby desirable to enhance the transduction efficiency in integrated electro-opto-mechanical resonators. For a resonator with loaded optical cavity linewidth $\kappa$, oscillating at a mechanical resonance frequency $\Omega_m$, the motion leads to predominant frequency modulation of the circulating light field \cite{ipj} when $\Omega_m \gg \kappa$ (resolved sideband regime). For transduction of signals with mechanical frequencies in the resolved sideband regime using a singlet WGM optical resonance, we see in Figure \ref{doubletboostillus}, that both the Stokes and anti-Stokes motional sidebands lie outside the optical cavity. The component of the intra-cavity energy at the mechanical resonance frequency is hence diminished, as compared to transduction of signals in the unresolved sideband regime. This leads to incomplete modulation, resulting in a low RF signal amplitude when the modulation is sensed using a photo-detector. To counter this inefficiency, we explore the possibility of exploiting Rayleigh scattering induced optical mode splitting in optical whispering gallery mode (WGM) resonators \cite{Gorodetsky} to transduce signals in the resolved sideband regime more efficiently, by using the optical mode doublet to boost the Stokes sideband. Back-scattering centers in the resonator cause the degeneracy of the clockwise (CW) and the counter clockwise propagating modes to be lifted and lead to splitting of the otherwise originally degenerate optical modes \cite{Gorodetsky, borselli, kippscatt}. The following section presents theoretical analysis of this transduction technique. Later we discuss experimental results based on our observations and demonstrate efficient modulation in a silicon coupled opto-mechanical resonator at frequencies up to 9.1GHz. Optical doublet resonances could also be engineered via Focused Ion Beam (FIB) engineering \cite{FIB} and this universal sense scheme could readily be extended to transduce mechanical motion at multi-GHz frequencies in piezo-opto-mechanical \cite{hongtang,cleland} and all optical transduced systems \cite{painter1,painter2}.

\section{Theoretical Formulation}

\begin{figure}[h]
\centerline{
\includegraphics[width=13cm]{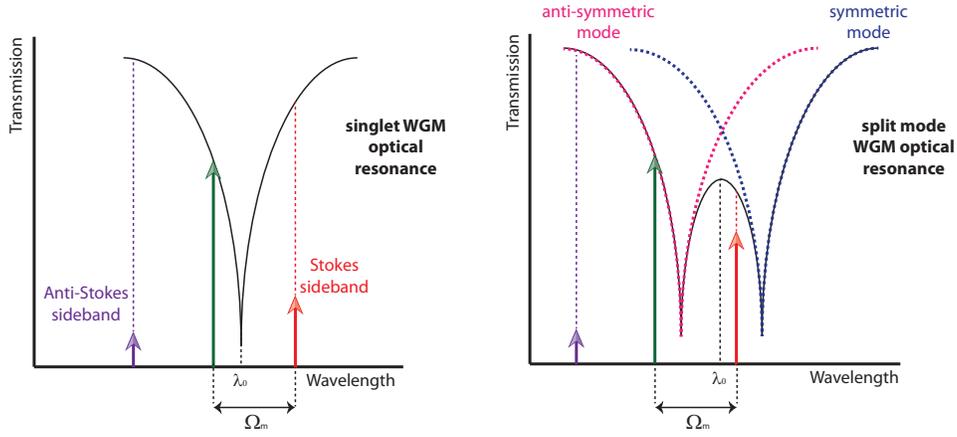}}
 \caption{Illustration of Rayleigh scattering induced optical mode splitting for enhancement of transduction efficiency of the optical sense scheme. The Stokes sideband amplitude is boosted by the presence of the second optical resonance as seen in the right panel.}
 \label{doubletboostillus}
\end{figure}

Consider the case of an opto-mechanical transducer sensed using modulation of CW laser light coupled to a back-scattering induced doublet optical resonance instead of a single optical WGM resonance as shown in Figure \ref{doubletboostillus}. The system under study comprises of coupled ring silicon acousto-optic modulator fabricated using the process flow described in \cite{detector}. The resonator is designed to have a compound radial expansion mode of vibration at 1.1GHz and higher order radial expansion modes at multiples of this frequency. Figure \ref{modesplitting} (a) shows a scanning electron micrograph (SEM) for this device. 

The coupling of the two modes is quantified in terms of a coupling quality factor $Q_u$. The transmission equation for the optical resonance in presence of mode splitting can be written as follows \cite{qlidoublet}:

\begin{equation}
\label{transdoublet}
T(\omega)=\left|1-\frac{1}{2Q_{ext}}\left(\frac{1}{j\left(\delta+\frac{1}{2Q_u}\right)+\frac{1}{2Q_{int}}+\frac{1}{2Q_{ext}}}+\frac{1}{j\left(\delta-\frac{1}{2Q_u}\right)+\frac{1}{2Q_{int}}+\frac{1}{2Q_{ext}}}\right)\right|^2
\end{equation}

%

Here we assume both optical modes to have the same intrinsic optical quality factor $Q_{int}$ and extrinsic coupling quality factor $Q_{ext}$. The detuning of the optical frequency $\omega$ from the optical resonance frequency $\omega_{opt}$ is defined as $\delta=\frac{\omega-\omega_{opt}}{\omega_{opt}}$. Following a coupled mode approach, the mean field amplitudes of the coupled CW and CCW modes may be written in terms of the input field, $\overline{s}$ as follows \cite{kippscatt} (see Figure \ref{modesplitting} (b)):

\begin{figure}[ht]
\centerline{
\includegraphics[width=13cm]{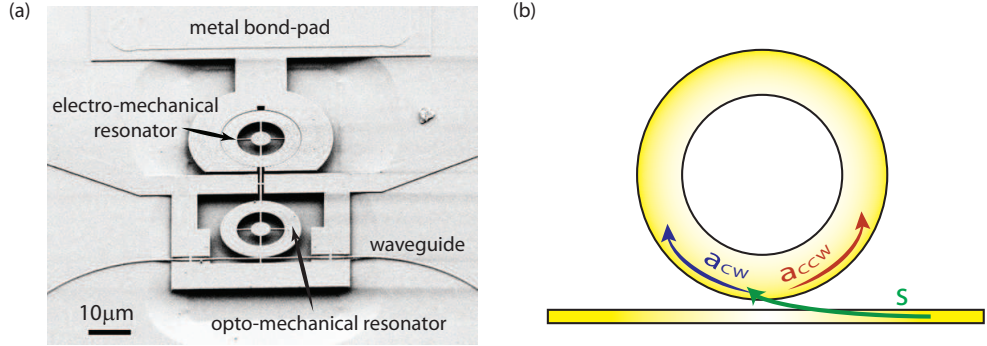}}
\caption{(a) Scanning electron micrograph (SEM) of the coupled silicon opto-mechanical resonator. The ring resonators have an inner radius of 5.7$\mu$m and an outer radius of 9.5$\mu$m. The thickness of the silicon device layer is 220nm. (b) Illustration of clockwise (CW) and counter-clockwise (CCW) propagating optical modes. The CW propagating mode is pumped by the input laser light field $s$.}
\label{modesplitting}
\end{figure}

\begin{equation}
\overline{a}_{CW} = \overline{s}\sqrt{\Gamma_{ext}}\frac{j\Delta-\frac{\Gamma_{tot}}{2}}{\Delta^2-\frac{\Gamma_{tot}^2}{4}-\frac{\gamma^2}{4}+j\Delta\Gamma_{tot}}
\label{aCW}
\end{equation}

\begin{equation}
\overline{a}_{CCW}=\frac{-j\frac{\gamma}{2}}{j\Delta-\frac{\Gamma_{tot}}{2}}\overline{a}_{CW}
\label{aCCW}
\end{equation}

It is reasonable to assume that only the clockwise mode is pumped by the input field. This is in turn coupled to the counterclockwise mode via the scattering rate $\gamma$. $\Gamma_{ext}$ is the decay rate associated with coupling of photons to the optical cavity and $\Gamma_{tot}$ is the loaded cavity photon decay rate. The detuning $\Delta$ is specified for the originally degenerate optical mode, with optical resonance frequency $\omega_{opt}$. We also assume that $\gamma$ and $\Gamma_{tot}$ are the same for both optical modes. The CW mode is coupled to the output field via $\overline{s}_{out}=\overline{s}-\overline{a}_{CW}\sqrt{\Gamma_{ext}}$. The coupling of the CW and CCW modes leads to formation of a mode doublet, which can be transformed into a pair of symmetric and antisymmetric modes \cite{b1b2}, $\widehat{b}_1=\frac{\overline{a}_{CW}+\overline{a}_{CCW}}{\sqrt{\Gamma_{ext}}}$ and $\widehat{b}_2=\frac{\overline{a}_{CW}-\overline{a}_{CCW}}{\sqrt{\Gamma_{ext}}}$, representing the lower and higher frequency modes respectively. The frequencies of these modes are $\omega_1=\omega_{opt}-\frac{\omega_{opt}}{2Q_u}$ and $\omega_2=\omega_{opt}+\frac{\omega_{opt}}{2Q_u}$ respectively. Thus, the detuning values of the input laser line with respect to these two modes are $\Delta_1=\Delta+\frac{\omega_{opt}}{2Q_u}$ and $\Delta_2=\Delta-\frac{\omega_{opt}}{2Q_u}$ respectively.

Following the derivations in \cite{b1b2, jesse}, we can write down the intra-cavity field values for $\widehat{b}_1$ and $\widehat{b}_2$ in presence of mechanical motion $x(t)=x_0sin(\Omega_mt)$, which corresponds to modulation index $\beta=\frac{x_0}{R}\frac{\omega_{opt}}{\Omega_m}$:

\begin{multline}
\widehat{b}_{1,intra}=\overline{s}\frac{\sqrt{\Gamma_{ext}}\left[j\Delta+\frac{\Gamma_{tot}}{2}\right]\left[j\left(\Delta+\frac{\gamma}{2}\right)+\frac{\Gamma_{tot}}{2}\right]}{\Delta^2-\frac{\Gamma_{tot}^2}{4}-\frac{\gamma^2}{4}+j\Delta\Gamma_{tot}}
\\\sum_{n=-\infty}^{+\infty}\frac{(-i)^nJ_n(\beta)}{\frac{\Gamma_{tot}}{2}+j\left(\Delta_1+n\Omega_m\right)}e^{j\left[\left(\omega_{opt}+n\Omega_m\right)t+\beta cos\left(\Omega_mt\right)\right]}
\label{b1}
\end{multline}

\begin{multline}
\widehat{b}_{2,intra}=\overline{s}\frac{\sqrt{\Gamma_{ext}}\left[j\Delta+\frac{\Gamma_{tot}}{2}\right]\left[j\left(\Delta+\frac{\gamma}{2}\right)+\frac{\Gamma_{tot}}{2}\right]}{\Delta^2-\frac{\Gamma_{tot}^2}{4}-\frac{\gamma^2}{4}+j\Delta\Gamma_{tot}}
\\\sum_{n=-\infty}^{+\infty}\frac{(-i)^nJ_n(\beta)}{\frac{\Gamma_{tot}}{2}+j\left(\Delta_2+n\Omega_m\right)}e^{j\left[\left(\omega_{opt}+n\Omega_m\right)t+\beta cos\left(\Omega_mt\right)\right]}
\label{b2}
\end{multline}

\begin{figure}[ht]
\centering
\includegraphics[width=13cm]{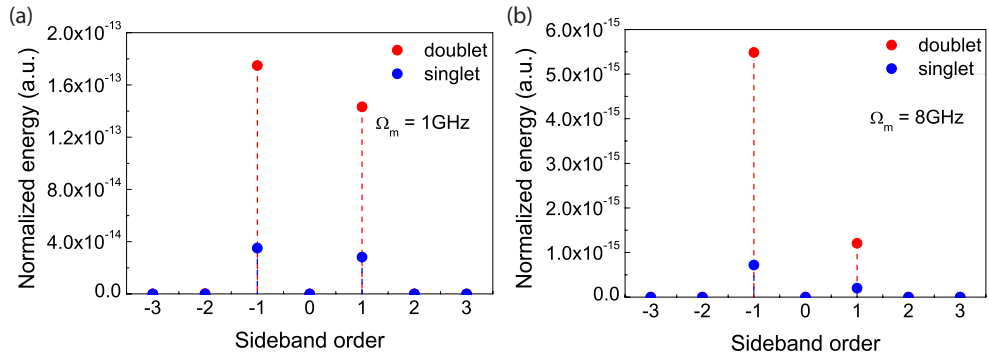}
 \caption{Comparison of sideband amplitudes (normalized to the laser power) for two mechanical modes at 1GHz and 8GHz in case of (a) singlet and (b) doublet resonances. We can clearly see the large boost in intra-cavity energy provided by the optical doublet for the Stokes sideband for the case where the mechanical frequency assumed is 8GHz. The pump laser line (sideband order = 0) is suppressed for easy visualization.}
 \label{doubletcomparisontheory}
\end{figure}

The total intra-cavity energy is given by $\left|\overline{a}_{CW}\right|^2+\left|\overline{a}_{CCW}\right|^2$. In Figure \ref{doubletcomparisontheory}, we plot the total intra-cavity energy in the case of an optical doublet, and compare it to $\left|\widehat{b}_{2,intra}\right|^2$ to get an idea of the net improvement contributed by presence of the other optical resonance, $\widehat{b}_1$. We assume $\Delta$ = 11GHz, $Q_{tot}$ = 70,000, $Q_{ext}$ = 50,000, $Q_u$ = 20,000, $\lambda_{opt}$ = 1,564nm, $x_0$ = 6.75pm, $R$ = 9.5µm. The frequency separation between the two optical modes in this case is 9.59GHz. We can clearly see the large boost in intra-cavity energy provided by the optical doublet for the Stokes sideband for the case where the mechanical frequency assumed is 8GHz.

\section{Experimental Characterization}

\begin{figure}[ht]
\centering
\includegraphics[width=13cm]{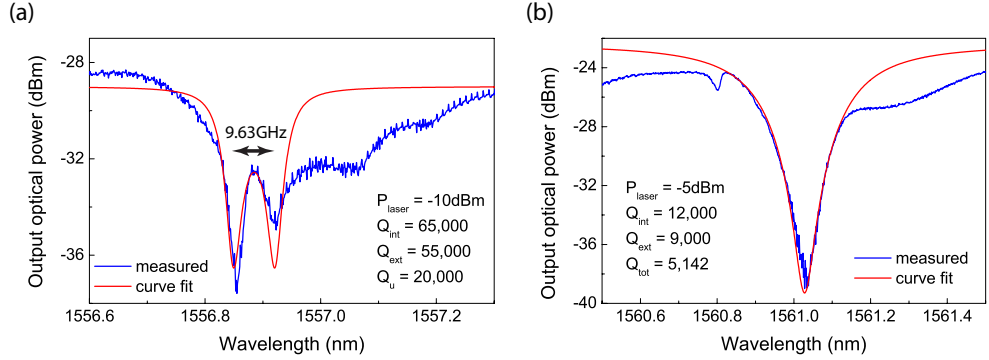}
 \caption{(a) A split optical resonance for the silicon coupled ring resonator. The frequency difference between the two resonances in the optical doublet is 9.63GHz. (b) A singlet optical resonance for the silicon coupled ring resonator with loaded optical quality factor 5,142.}
 \label{opticaldoubletsinglet}
\end{figure}

We identify a split optical resonance and a singlet optical resonance in the coupled opto-mechanical resonator. Figure \ref{opticaldoubletsinglet} shows transmission spectra for these optical modes. To test the validity of this theory, we study the electro-mechanical transmission using the doublet resonance at 1,556.9nm and the singlet resonance at 1,561nm. The frequency difference between the two resonances in the optical doublet is 9.63GHz.

\begin{figure}[ht]
\centering
\includegraphics[width=12cm]{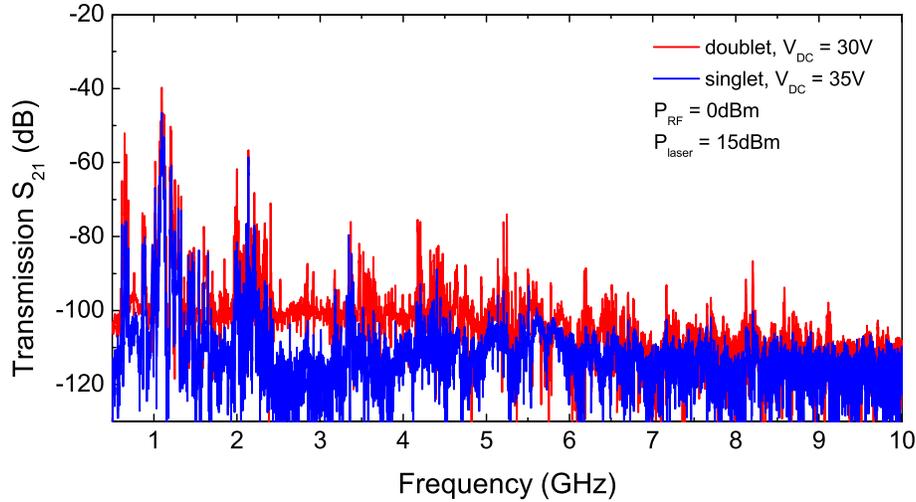}
 \caption{Electromechanical transmission spectrum for the coupled ring resonator. The signals at higher frequencies have larger amplitudes when we employ the optical doublet for sensing motion.}
 \label{doubletwidesweep}
\end{figure}

To study the mechanical transmission spectrum, we apply a DC voltage bias to the electrodes and use an Agilent N5230 PNA-L network analyzer and a NewFocus 1544-A photodetector and follow the experiment detailed in \cite{ipj}. The resonator-electrode gap is 130nm. As we clearly see in Figure \ref{doubletwidesweep}, the doublet resonance boosts the transduction of signals at higher frequencies (5GHz - 10GHz) owing to the second optical resonance. The enhancement is most pronounced at 5.25GHz and 8.2GHz, where the insertion loss improves by 25dB and 14dB respectively. Also, it enables us to observe signals at frequencies all the way up to 9.1GHz and 9.8GHz, which are not possible using the singlet optical resonance.

\begin{figure}[ht]
\centering
\includegraphics[width=13cm]{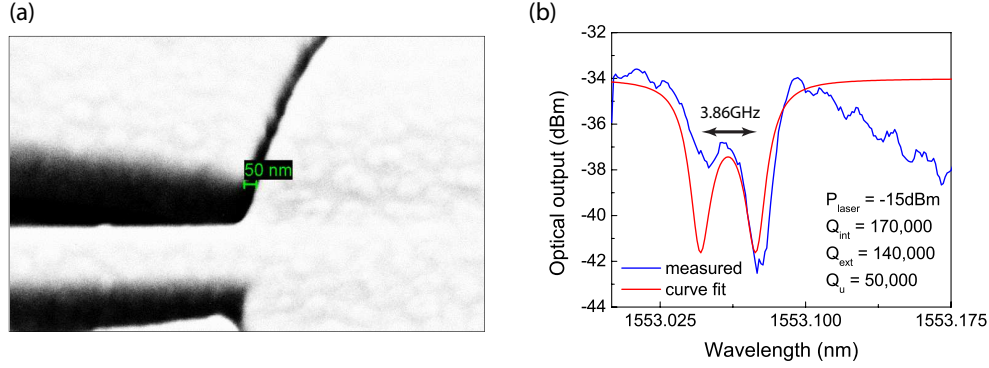}
 \caption{(a) SEM showing the reduced resonator-electrode gap via ALD. (b) A split optical resonance with a frequency splitting of 3.86GHz between the modes.}
 \label{50nm_doubletopt}
\end{figure}

Next we examine this coupled resonator system following a partial gap process flow \cite{T13} to improve the electrostatic drive transduction efficiency. This process is used to reduce the resonator electrode gap from 130nm to 50nm via atomic layer deposition (ALD) of alumina (Al$_2$O$_3$). Figure \ref{50nm_doubletopt} (a) shows an SEM of this reduced gap. We identify an optical doublet resonance in this device with a frequency splitting of 3.86GHz (see Figure \ref{50nm_doubletopt} (b)).

\begin{figure}[ht]
\centering
\includegraphics[width=12cm]{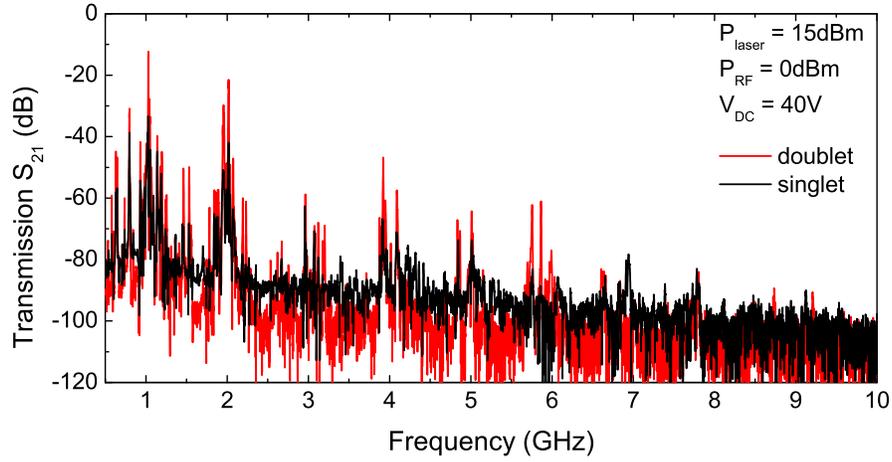}
 \caption{Electromechanical transmission spectra for a 50nm gap resonator highlighting the efficacy of combining the doublet-based sensing scheme with a partial gap transduced drive scheme. The signal strength for the fourth order radial mode at 4GHz shows an improvement of 32dB over the signal recorded at 4.4GHz in Figure \ref{doubletwidesweep} pre-ALD using a singlet resonance.}
 \label{50nm_doublet}
\end{figure}

Figure \ref{50nm_doublet} shows a comparison of the electromechanical transmission spectra recorded in this device using the doublet resonance, in comparison to transduction using a singlet resonance with loaded optical quality factor $\approx$60,000.

Comparing this to Figure \ref{doubletwidesweep}, it is easy to observe the larger signal strengths recorded using a doublet resonance in combination with partial gap transduction. The resonator thickness prior to ALD is 220nm. Depositing 40nm ALD alumina on all surfaces of the resonator predominantly changes the thickness, thereby changing the effective mass of the mechanical mode $m_{eff}$ and hence the mechanical resonance frequency $\left(\Omega_{mech} \propto \sqrt{\frac{1}{m_{eff}}}\right)$. Thus, the resonance frequencies are roughly expected to lower by a factor of $\sqrt{\frac{\rho_{Si}*220}{\rho_{Si}*220+\rho_{Al_2O_3}*80}} \approx 0.83$. We experimentally observe a shift in the mechanical resonance frequencies for the radial mode family from 1.1GHz to 1GHz for the fundamental mode, from 4.4GHz to 4GHz for the fourth order mode and so on for each mode order. Notable signal strength enhancement is observed for the fourth order mode (32dB increase from -78dB to -46dB) and the sixth order mode (28dB increase from -90dB to -62dB) compared to a singlet resonance without ALD. The mass loading on account of ALD also lowers the mechanical quuality factors (980 for the fourth order radial mode post-ALD, 3,500 pre-ALD). The amplitude of motion is directly proportional to the mechanical quality factor $Q_{mech}$ and varies as inverse-squared power of the resonator-electrode gap $g$ and the frequency $\Omega_{mech}$ $\left(U_0\propto \frac{Q_{mech}}{\Omega_{mech}g^2}\right)$ \cite{FCS11}. The lowering of the mechanical quality factor counters the enhancement due to reduced gap and hence no improvement in signal strength is noticed at frequencies beyond 7GHz. The signal strength is largest at the mechanical resonance frequencies whose values are close to the frequency split in the optical resonances, on account of maximum overlap of the Stokes sideband with the adjacent optical mode. This frequency split could be engineered via changing the resonator-waveguide gap using electrostatic transduction \cite{mingwu} thus providing a tunable mode selection mechanism.

\section{Conclusion}
This paper presents a novel transduction scheme utilizing Rayleigh scattering in opto-mechanical resonators to enhance displacement sensitivity at multi-GHz frequencies in the resolved sideband regime. A combination of partial gap transduction and Rayleigh scattering based sensing was used to improve transduction efficiency in a coupled silicon electro-opto-mechanical resonator by 32dB at 4GHz and 28dB at 6GHz. The universal nature of this sense scheme could also potentially push other transduction mechanisms such as piezo-opto-mechanical and all optical schemes to frequencies in the microwave X-band, thereby adding to the vast variety of experiments that could be realized using opto-mechanics.

\section{Acknowledgments}

This work was supported by the DARPA/MTO's ORCHID program. The resonators were fabricated at the Cornell NanoScale Science and Technology Facility, which is supported by the National Science Foundation.

\end{document}